\documentclass{aa}  
\usepackage{graphicx}
\usepackage{txfonts}
\usepackage{multirow}%
\usepackage{amsmath,amssymb,amsfonts}%

\usepackage{amsthm}%
\usepackage{mathrsfs}%
\usepackage[title]{appendix}%
\usepackage{xcolor}%
\usepackage{textcomp}%
\usepackage{manyfoot}%
\usepackage{algorithm}%
\usepackage{algorithmicx}%
\usepackage{algpseudocode}%
\usepackage{listings}%
\usepackage{subcaption}
\usepackage{threeparttable}
\usepackage{tabularx}
\usepackage{booktabs}%
\usepackage{hyperref}

\begin{document}

   \title{Reassessing planetary composition: Evidence of rock-dominated envelopes in Uranus and Neptune}
\titlerunning{Evidence of rock-dominated envelopes in Uranus and Neptune}

   \author{Vanesa Ramirez
          \inst{1}
          \and
          Yamila Miguel
          \inst{1,2}
          \and
          Saburo Howard
          \inst{3}
          }

   \institute{Leiden Observatory, Leiden University,
              PO Box 9513, 2300 RA, Leiden, The Netherlands\\
              \email{ramirez@strw.leidenuniv.nl}
         \and
             SRON Netherlands Institute for Space Research, Niels Bohrweg 4, 2333 CA, Leiden, The Netherlands
         \and
             Institut für Astrophysik, Universität Zürich, Winterthurerstr. 190, 8057, Zürich, Switzerland
             }

   \date{Received January 23 2026 / Accepted April 10 2026}

  \abstract
  {Although Uranus and Neptune are commonly classified as ice giants, their exact compositions remain poorly constrained. Recent studies of outer Solar System bodies challenge the traditional view that these planets are primarily ice-dominated, suggesting that refractory material plays a more significant role. Determining the proportions of ice and rock within Uranus and Neptune is essential for understanding their formation and the evolutionary history of the Solar System. In this work we computed interior structure models for both planets and explored, within a Bayesian framework, the range of compositions that satisfy the available observational constraints. We quantified the resulting ice and rock fractions and analyzed their impact on the inferred internal structure. Our results suggest that the envelopes of both Uranus and Neptune are systematically enriched in refractory material, with median rock fractions of approximately 60\% within the heavy-element component, similar to Pluto, Kuiper belt objects, and comets. In contrast, the deep interiors of the two planets exhibit distinct compositions: Neptune is best fit by relatively rock-rich mantles (median rock fraction of $\sim$55\%), whereas Uranus is inferred to have more ice-rich mantles (median rock fraction of $\sim$41\%), consistent with a more strongly stratified structure. These results point to compositional differences between Uranus and Neptune that may reflect divergent formation and evolutionary pathways.}


   \maketitle

\section{Introduction}

Understanding the interior compositions of Uranus and Neptune is essential for advancing our knowledge of planetary formation and evolution, both within the Solar System and beyond. These two planets serve as key reference points for interpreting exoplanets of similar sizes, particularly those in the Neptune and sub-Neptune category, which account for nearly 40\% of the confirmed exoplanet population\footnote{Based on data from the NASA Exoplanet Archive: \url{https://exoplanetarchive.ipac.caltech.edu}}. Consequently, they provide a unique opportunity to refine planetary structure models and characterize analogous worlds outside the Solar System.

Although Uranus and Neptune have historically been labeled as ice giants, the constraints on their internal structures are highly nonunique, and their actual compositions remain uncertain and depend heavily on the assumptions underlying their models, as well as the materials used to represent heavy elements (see \citealt{Helled2020a} for an in-depth discussion). Measurements suggest relatively low deuterium-to-hydrogen (D/H) ratios in their atmospheres, which implies that, if the planets were once fully mixed, they now contain substantial amounts of rock in their interiors \citep{Feuchtgruber2013, Teanby2020}. Additional evidence from outer Solar System bodies supports this hypothesis: \cite{McKinnon2017} find a rock fraction of about 70\% for Pluto, while \cite{Bierson2019} report consistent rock mass fractions of 70\% for 15 Kuiper belt objects. Moreover, studies of comets indicate that their refractory-to-ice mass ratios range from 3 to 6 \citep{Rotundi2015, Fulle2016, Fulle2017, Fulle2019, Choukroun2020}. These findings suggest that Uranus and Neptune also contain significant rock components \citep{Teanby2020}, and that models that do not take into consideration the presence of rocks in their envelopes may oversimplify their internal structure.

Constraining the ice-to-rock fractions in Uranus and Neptune is critical for understanding their formation and evolution, as their formation location directly influences their accretion processes and resulting compositions \citep{Izidoro2015}. Furthermore, insights into the distribution of ice and rock within these planets could also help refine models of analogous exoplanets, in which rocks are often considered to reside only in the core. 

We addressed these questions by analyzing the internal structures of Uranus and Neptune and determining their ice and rock fractions. The structure of the paper is as follows. Section \ref{sec:methods} describes the interior structure models, the Bayesian inference framework, and the observational constraints adopted. In Sect. \ref{sec:results} we present our inferred heavy-element distributions, including ice and rock fractions, and their impact on the planetary structure. Section \ref{sec:discussion} discusses the implications of these results for the formation and evolution of Uranus and Neptune. In Sect. \ref{sec:conclusions} we conclude with a summary of our main findings.

\section{Methods}\label{sec:methods}

\newcolumntype{Y}{>{\centering\arraybackslash}X}
\begin{table*}[t]
\captionsetup{justification=raggedright,singlelinecheck=false,width=\textwidth}
\caption{Observational constraints for Uranus and Neptune used in this work.}
\label{table:prop}
\centering
\begin{threeparttable}
\begin{tabularx}{0.8\textwidth}{lYY}
\toprule
\toprule
& Uranus & Neptune \\
\midrule
Planet mass (kg $\times 10^{25}$) & 8.68$^{\rm a}$ & 10.241$^{\rm a}$ \\
Measured equatorial radius at 1 bar (km) & 25559$^{\rm b}$ & 24766$^{\rm c}$ \\
Temperature at 1 bar (K) & 76 $\pm$ 2$^{\rm b}$ & 72 $\pm$ 2$^{\rm c}$ \\
Revised solid-body rotation period (s) & 59664$^{\rm d}$ & 62849$^{\rm d}$ \\
Quadrupole gravitational harmonic $J_2 \times 10^{-2}$ & 0.35107(7)$^{\rm e}$ & 0.35294(45)$^{\rm f}$ \\
Octopole gravitational harmonic $J_4 \times 10^{-4}$ & -0.342(13)$^{\rm e}$ & -0.358(29)$^{\rm f}$ \\
\bottomrule
\end{tabularx}
\begin{tablenotes}
\item $^{\rm a}$ via \href{https://ssd.jpl.nasa.gov/horizons/app.html#/}{JPL Horizons}
\item $^{\rm b}$ \cite{Lindal1987}
\item $^{\rm c}$ \cite{Lindal1992}
\item $^{\rm d}$ \cite{Helled2010}
\item $^{\rm e}$ \cite{Jacobson2014}
\item $^{\rm f}$ \cite{Jacobson2009}
\end{tablenotes}
\end{threeparttable}
\end{table*}

\subsection{Calculation of the Uranus and Neptune interior models}

\begin{figure}[!t]
\centering 
\includegraphics[width=
0.48\textwidth]{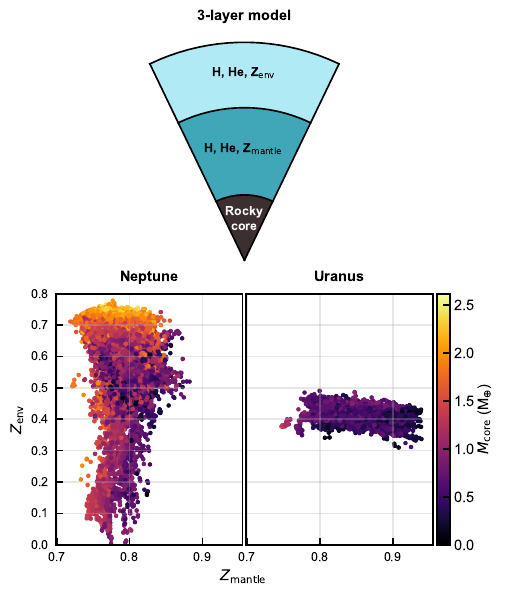}
    \caption{Details of the interior models of Uranus and Neptune. Top: Illustration of the three-layer internal structure model assumed. Both the envelope and mantle contain hydrogen, helium, ices, and rocks, with the mantle expected to have a higher concentration of ices and rocks. The core is composed entirely of rocks. Bottom: Envelope metallicity versus mantle metallicity for interior models of Uranus and Neptune. Each dot represents a model that fits both the gravitational data and radius of the respective planet. The color scale indicates the core mass.}
    \label{fig:zatmversuszdeep}
\end{figure}

We constructed models assuming a three-layer structure: envelope, mantle, and a fully rocky core (see Fig.~\ref{fig:zatmversuszdeep}, left panel). This approach has been employed in previous studies \citep{Bailey2021, Scheibe2019, Nettelmann2013}, all of which assume that rocks are confined solely to the core \footnote{\cite{Bailey2021} also explore an alternative two-layer model, where there is no separate rocky core; instead, rocks are fully mixed with ices in the mantle, while the outer envelope remains rock-free.}. In contrast to these works, we allowed both ices and rocks to be present in the planetary envelope and mantle to determine the ice-to-rock fractions that fit the data. Following \citet{Cano2024}, we restricted the presence of rocks to regions where the temperature exceeds a threshold consistent with silicate condensation. We adopted a condensation threshold of 1500~K as a proxy for the silicate cloud base, since silicates are the dominant rocky component (70\% SiO$_2$) and form the basis of the metal equation of state (EOS) considered here \citep{Lyon1992}. Furthermore, this choice avoids numerical instabilities when the condensation boundary coincides with the envelope–mantle transition. Because condensation curves depend strongly on composition and mixing, this value should be regarded as a conservative lower limit for the onset of rock condensation in the interiors of these planets.

To fit the measured gravity data for both planets, a sharp transition in composition from the hydrogen-rich outer envelope to the deeper regions containing heavier elements appears necessary. The physical cause of this discontinuity is not yet fully understood, but hydrogen-water immiscibility is a possible explanation, and this transition must occur at shallow depths ($\sim 0.7~R_{\rm planet}$; \citealt{Bailey2021, Bergermann2024, Cano2024,Gupta2025}). Although hydrogen-water immiscibility has been demonstrated at conditions relevant to the Earth's mantle \citep{Bali2013, Vlasov2023}, there remains a lack of experimental data for the extreme pressures and temperatures inside the ice giants. Given these uncertainties, we treated the depth of the compositional transition as a model parameter and explored a wide range of possibilities.

\begin{table*}[!t]
\centering
\begin{threeparttable}
\captionsetup{justification=raggedright,singlelinecheck=false,width=\textwidth}
\caption{Priors for model parameters used in both the Uranus and Neptune simulations.}
\label{table:priors}
\begin{tabularx}{0.8\textwidth}{
    l
    l
    >{\centering\arraybackslash}X
    >{\centering\arraybackslash}X
    >{\centering\arraybackslash}X
    >{\centering\arraybackslash}X
}
\toprule
\toprule
Parameter & Distribution & Lower bound & Upper bound & $\mu$ & $\sigma$ \\
\midrule
$M_{\rm core}$ (M$_{\rm p}$)    & Uniform & 0 & 0.2 (0.1) & \textendash & \textendash \\
$Z_{\rm env,\; ice}$            & Uniform & 0 & 0.7       & \textendash & \textendash \\
$Z_{\rm env,\; rock}$           & Uniform & 0 & 0.7       & \textendash & \textendash \\
$Z_{\rm mantle,\; ice}$         & Uniform & 0 & 0.99      & \textendash & \textendash \\
$Z_{\rm mantle,\; rock}$        & Uniform & 0 & 0.99      & \textendash & \textendash \\
$P_{1-2}$ (Mbar)                & Uniform & 0.01 & 0.4 & \textendash & \textendash \\
$T_{\rm 1bar}$ (K)              & Normal  & 50 & 100 & 72 (76) & 4 \\
\bottomrule
\end{tabularx}
\end{threeparttable}

\vspace{0.5em}

\noindent\begin{minipage}{\textwidth}
\footnotesize
\textbf{Notes.} For normally distributed parameters, $\mu$ and $\sigma$ denote the mean and standard deviation, respectively. Most priors are identical for the two planets, except for $M_{\rm core}$ and $T_{\rm 1bar}$. 
The differing priors for $M_{\rm core}$ are to ensure convergence in Uranus models, while the variation in $T_{\rm 1bar}$ is based on observational data. 
For parameters with different priors, values in parentheses refer to Uranus, and those outside the parentheses correspond to Neptune.
\end{minipage}
\end{table*}

While this study employs three-layer models with distinct regions for the core, mantle, and envelope, the actual interior structures of Uranus and Neptune may be more complex. In particular, recent studies suggest that compositional gradients are required to explain Uranus’s anomalously low luminosity, which is otherwise difficult to reconcile with a fully convective interior \citep{Scheibe2019, Vazan2020, Arevalo2025}. Several recent works explore interior models with continuous compositional gradients rather than sharp boundaries for Uranus and Neptune (e.g., \citealt{Neuenschwander2024, Malamud2024, Morf2024, Lin2025, Morf2025, Wirth2026}). Such gradients imply a departure from the discrete layering assumed in our models, resulting in partial mixing of heavy elements across regions. However, the physical processes responsible for the formation and long-term stability of these gradients remain poorly understood and require further investigation.

Conversely, other studies provide support for classical three-layer models. For instance, \citet{Cano2024} suggest that a layered structure can arise from a combination of primordial mass accretion and subsequent phase separation between hydrogen and water \citep[see also][]{Howard2025}. Similarly, \citet{Bailey2021} argue that models relying on substantial density gradients rather than discrete layers struggle to satisfy the requirement for a well-mixed region of sufficient radial extent to sustain dynamo-generated magnetic fields. Such discrete layers are likely only stable in the presence of immiscible phases, a scenario that has recently gained support from ab initio calculations \citep{Militzer2024}. While our current approach assumes distinct layers, we acknowledge that compositional gradients remain a plausible alternative and an important avenue for future work.

The observational data used for the modeling include the gravitational harmonics $J_2$ and $J_4$ as well as the planet's radius \citep{Lindal1987, Lindal1992}. Direct measurements of Uranus and Neptune were obtained from brief flybys by NASA’s Voyager 2 mission in the 1980s, providing estimations for mass, radius, bulk density, and gravitational field \citep{Tyler1986, French1988, Tyler1989}. Subsequent observations of the planets' satellites have refined these gravity field estimates, offering updated assessments of shape and solid-body rotation periods \citep{Jacobson2009, Jacobson2014, French2024}, which we incorporated into our models. We adopted the solid-body rotation periods calculated by \cite{Helled2010}. The main physical parameters used in our models for Uranus and Neptune are listed in Table \ref{table:prop}. Regarding envelope composition, the only constraint imposed is a constant hydrogen-to-helium ratio, set to the protosolar value of 0.277.

An accurate determination of the internal structure of the ice giants is tightly linked to the reliability of the EOSs used to describe materials under the extreme conditions present in their interiors. For hydrogen, we adopted the MH13-H EOS \citep{Militzer2013, Miguel2016, Miguel2018}, and for helium, the SCVH95-He EOS \citep{Saumon1995}. For ices, we assumed pure water, as in previous studies \citep{Helled2011, Nettelmann2013, Scheibe2019, Scheibe2021, Bailey2021}. We implemented a state-of-the-art water EOS calculated by \citet{Scheibe2019, Scheibe2021} and based on the works of \citet{Mazevet2019} and \citet{French2009}, hereafter referred to as the REOS. Rocks were modeled using the SESAME EOS \citep{Lyon1992} for a dry sand mixture whose main components by weight are SiO$_2$ (70\%), Al$_2$O$_3$ (8.22\%), Fe$_2$O$_3$ (4.53\%), and CaO (4.25\%). Given that the interiors of Uranus and Neptune are expected to be highly enriched in heavy elements, the choice of EOS for both ices and rocks is likely to play a critical role in determining their inferred internal structures. To assess the sensitivity of our results to these choices, we also ran comparative models using alternative EOSs: the SESAME water EOS \citep{Lyon1992} for the ice component, and QEOS for SiO$_2$ \citep{More1988, Vazan2013} for the rocky component.

We computed each interior structure model with the code CEPAM \citep{cepam}, which is widely used to study the giant planets in our Solar System \citep{Miguel2016, Miguel2022, Howard2023_int}. CEPAM solves the interior structure equations and estimates the gravitational harmonics of the resulting density structure using the Theory of Figures \citep{Zharkov1978} implemented up to the seventh order \citep{Nettelmann2021}. The modeled structure and gravitational harmonics are then compared against observational constraints $J_2$ and $J_4$ and the planetary radius, to ensure that it fits the available data.

\subsection{Bayesian framework and parameter space exploration}

To address the degeneracies inherent in planetary interior modeling, where multiple solutions can fit the observed data, we employed a Bayesian framework that enables the efficient exploration of model parameters. Building on the Bayesian models developed by \cite{Bazot2012} and \cite{Miguel2022}, we adapted this framework specifically for the ice giants and perform Markov chain Monte Carlo (MCMC) simulations. This approach enabled us to investigate key regions of the parameter space by evaluating hundreds of thousands of potential solutions (i.e., models that satisfy the observational constraints for Uranus and Neptune) at a feasible computational cost.

The MCMC model parameters include: core mass ($M_{\rm core}$), the transition pressure between the envelope and the mantle ($P_{1-2}$), the mass fraction of heavy elements, referred to as metallicity, in the envelope ($Z_{\rm env}$) and the mantle ($Z_{\rm mantle}$), and the mass fractions of ices and rocks in both the envelope ($Z_{\rm env,\; ice}$, $Z_{\rm env,\; rock}$) and the mantle ($Z_{\rm mantle,\; ice}$, $Z_{\rm mantle,\; rock}$). Additionally, the temperature at 1 bar ($T_{\rm 1bar}$) is considered as the outer boundary. The prior distributions for these parameters are outlined in Table \ref{table:priors}. We selected these priors to explore a broad range of values for each parameter while ensuring model convergence. We verified that widening the Uranus core-mass prior does not qualitatively alter the inferred ice and rock fractions, but substantially degrades MCMC convergence, with model solutions failing to reproduce $J_4$.

\section{Results}\label{sec:results}

\subsection{Planetary metallicities}

Figure~\ref{fig:zatmversuszdeep} presents the posterior distributions of envelope metallicity ($Z_{\rm env}$) and mantle metallicity ($Z_{\rm mantle}$) for Uranus and Neptune interior models. We note that these values represent effective global metallicities: the rock contribution in each layer is scaled by the mass fraction of the region where rocks are present, as determined by the silicate cloud base (see Sect. \ref{sec:methods} for details). Our results reveal that both planets possess significant mantle enrichment, with $Z_{\rm mantle}$ consistently exceeding 0.7 across all accepted models. A distinct dichotomy exists between the two planets. As shown in the right panel, Uranus exhibits a tightly constrained solution space, characterized by a lower envelope metallicity of $Z_{\rm env} = 0.410^{+0.016}_{-0.017}$ and a highly enriched mantle ($Z_{\rm mantle} = 0.860^{+0.026}_{-0.031}$). The darker hues of the cluster indicate that these solutions consistently favor a small central core, with a mass of $M_{\rm core} = 0.549^{+0.175}_{-0.149}\, M_{\oplus}$. Conversely, Neptune displays a vertically elongated distribution, reflecting a broader degeneracy in its internal structure. While the envelope metallicity is higher on average ($Z_{\rm env} = 0.602^{+0.117}_{-0.173}$), the values span a wide range, extending to lower metallicities that partially overlap with the regime inferred for Uranus. Neptune’s mantle metallicity ($Z_{\rm mantle} = 0.789^{+0.019}_{-0.021}$) is slightly lower than that of Uranus, while its core mass is larger and significantly less well constrained ($M_{\rm core} = 1.148^{+0.683}_{-0.417}\, M_{\oplus}$). The color mapping highlights a strong correlation for Neptune: solutions with high envelope metallicities require massive cores (yellow and orange regions), whereas solutions with lower $Z_{\rm env}$ correspond to smaller cores. This comparative lack of constraint is attributable to the larger uncertainties in Neptune's gravity harmonics ($J_2$, $J_4$) compared to the more precise measurements available for Uranus. Corner plots detailing the full parameter space are provided in Appendix \ref{secA1}.

In agreement with previous studies \citep{Nettelmann2013, Bailey2021}, our models confirm a highly enriched deep interior for both planets, consistent with the presence of a substantial heavy-element mantle. Our results also reproduce the Uranus–Neptune compositional dichotomy reported in those works, in which Neptune generally requires a higher concentration of heavy elements in its outer layers to satisfy the observed gravity field. However, our Bayesian retrieval indicates a systematically higher level of enrichment in the planetary envelope. While \cite{Nettelmann2013} found that Uranus could be fit with a relatively ``clean'' H--He envelope ($Z_{\rm env} \le 0.08$), our results suggest a more enriched outer layer ($Z_{\rm env} > 0.3$). Similarly, for Neptune, our median value of $Z_{\rm env} \sim 0.6$ lies at the upper end of the range previously explored. A primary distinction between our approach and previous studies is that we allowed for the presence of rocks not only in the core but also in the envelope and mantle. Considering the observed rock-rich compositions of outer Solar System bodies and atmospheric measurements of D/H and CO for Uranus and Neptune (see \citealt{Teanby2020} for an in-depth discussion), we argue that including rocks in the envelope and mantle provides a more realistic representation of their interiors (see Appendix \ref{secA0} for a test model in which rocks are confined to the core). Additionally, our choice of EOSs for the heavy elements (both ice and rock) has a substantial impact on the outcomes, which we explore in detail in Sect. \ref{sec:metals_eos}.

\subsection{Ice and rock fractions}

\begin{figure}[!t]
\centering 
\includegraphics[width=0.5\textwidth]{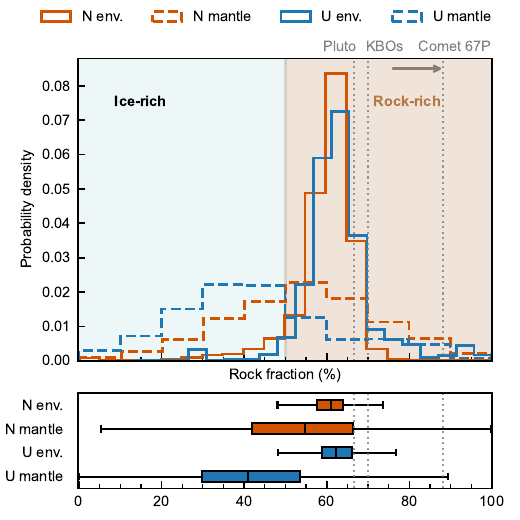}
\caption{Rock fractions in the ice giants. The distributions of rock fractions derived from our interior models are plotted for Neptune (red lines) and Uranus (blue lines). Top: Probability density of rock fractions for the envelope (solid lines) and mantle (dashed lines) of each planet. For comparison, rock fractions for Pluto \citep{McKinnon2017}, Kuiper belt objects (KBOs; based on a sample of 15 objects; \citealt{Bierson2019}), and the range of values reported for comet 67P/Churyumov--Gerasimenko \citep{Rotundi2015, Fulle2016, Fulle2017, Fulle2019} are also plotted, as vertical dotted gray lines. Note that the rock fraction is defined as the mass mixing ratio of rocks normalized by the total metallicity (rocks + ices) and therefore does not represent the total mass fraction of the layer. Bottom: Box plots that summarize the distribution of rock fractions in the envelope and mantle for both planets. Each box plot illustrates the interquartile range (IQR; spanning the 25th to 75th percentiles, represented by the width of the box), the median (black center line), and the data spread (whiskers extending to values within 1.5 times the IQR). Our results suggest that the envelopes of Uranus and Neptune are rock-dominated, with rock fractions consistent with those of outer Solar System objects.}
\label{fig:rock_fraction}
\end{figure}

While Uranus and Neptune have traditionally been classified as ice giants, their internal compositions remain poorly constrained, and the possibility that they may instead be better described as rock giants remains open \citep{Helled2011, Teanby2020, Neuenschwander2024, Morf2024, Morf2025}. To assess whether these planets are predominantly ice-rich or whether rocks constitute a major component of their interiors, we computed the relative fractions of ices and rocks for all accepted model solutions. Figure~\ref{fig:rock_fraction} shows the resulting distributions of rock fraction for both planets, separately for their envelopes and mantles. Here, the rock fraction is defined as the mass mixing ratio of rocks in a given layer, normalized by the total metallicity of that layer (i.e., the sum of the mass mixing ratios of rocks and ices).

The plot is divided into two regimes: an ice-rich region, where the rock fraction is below 50\%, and a rock-rich region, where it exceeds 50\%. For comparison, we also indicate rock fractions for Pluto, several Kuiper belt objects, and comet 67P/Churyumov–Gerasimenko, as reported in previous studies \citep{McKinnon2017, Bierson2019, Rotundi2015, Fulle2016, Fulle2017, Fulle2019}. All of these bodies fall within the rock-rich regime. The rock fractions inferred for the planetary envelopes are significantly more tightly constrained than those for the mantles, consistent with the fact that the gravity harmonics primarily probe the outer regions of the planets. Notably, the envelopes of both Uranus and Neptune are inferred to be rock-dominated, with 97\% and 92\% of our models, respectively, lying in the rock-rich regime, and with rock fraction values comparable to those of other outer Solar System bodies. In contrast, the mantle compositions exhibit a much broader range of solutions and are therefore substantially less well constrained. Despite this degeneracy, a clear difference emerges between the two planets: Uranus’s mantle solutions peak in the ice-rich regime, whereas Neptune’s mantle tends toward more rock-rich solutions.

\subsection{Abundances of key interior components}

\begin{figure}[!t]
\centering 
\includegraphics[width=0.5\textwidth]{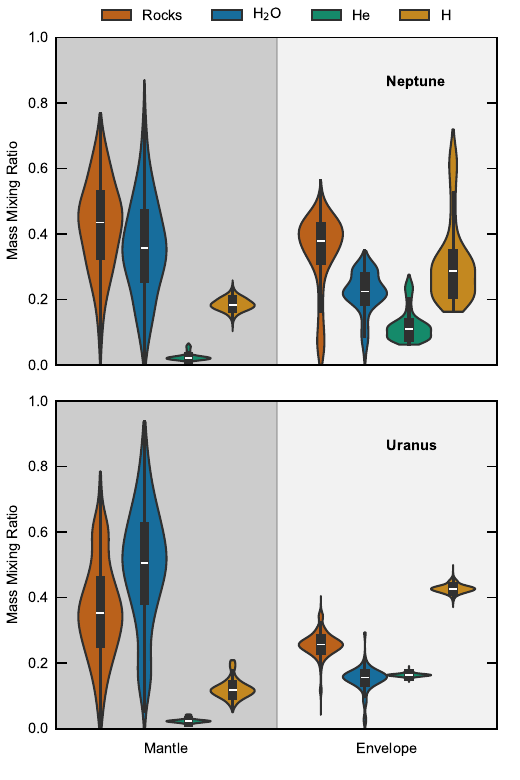}
\caption{Elemental abundances in the interiors of Uranus and Neptune. Violin plots depict the distribution of hydrogen, helium, water, and rock abundances across our interior solutions for the ice giants. Top: Neptune. Bottom: Uranus. The gray-shaded region on the left represents the mantle, while the lighter region on the right represents the envelope. Each violin plot combines the probability density (indicated by the width of the plot) with central abundance tendencies. The white horizontal line indicates the median, the thick black bar shows the IQR, and the thin black whiskers extend to the range of values within 1.5 times the IQR.}
\label{fig:abundances}
\end{figure}

Figure~\ref{fig:abundances} presents the posterior distributions of elemental mass mixing ratios within the interiors of Neptune (top panel) and Uranus (bottom panel). The violins illustrate the abundances of the four primary components considered in our modeling: hydrogen (H), helium (He), water (H$_2$O), and rocks (silicates), separated into the envelope (left panels) and mantle (right panels). The core is not depicted, as it is assumed to be composed entirely of rock.

Significant compositional differences are evident between the envelopes of the two planets. First, Uranus exhibits a systematically higher hydrogen abundance in its envelope compared to Neptune, a finding consistent with its lower overall metallicity (as shown in Fig. \ref{fig:zatmversuszdeep}) and lower mean density. Conversely, Neptune's envelope is characterized by a notably higher concentration of rocks. The distributions also reveal a distinct rock-to-water imbalance within the envelopes of both planets, where rocks are strongly favored over water. Furthermore, it is evident that the elemental abundances in Uranus's envelope are significantly better constrained than those of Neptune, a result consistent with the tighter constraints imposed by the available gravity data.

A clear compositional distinction exists between the layers: the envelopes are richer in hydrogen, whereas the mantles contain a major fraction of heavy elements (water and rocks). It is important to note that the uncertainties associated with the mantle composition are considerably larger than those for the envelope, particularly regarding the heavy-element ratios. This disparity is a direct consequence of the radial sensitivity of the gravitational harmonics ($J_n$); their contributions peak in the outer layers, thereby placing tighter constraints on the envelope while leaving the deep interior more degenerate \citep[see Fig. 2 of][]{Miguel2023}. Consequently, the large uncertainties in the measured $J_2$ and $J_4$ moments for the ice giants fundamentally limit our ability to resolve the deep internal structure, underscoring the critical need for a dedicated orbiter mission to these systems \citep[see also][]{Fletcher2022, Mousis2022, Movshovitz2022, Girija2023, Mandt2023, Mousis2024, Morf2024, Morf2025, Lin2025}.

\subsection{Impact of ice and rock abundances on the planet's pressure-temperature profile}

\begin{figure}[!t]
    \vspace{-25pt}
    \centering
        \includegraphics[width=0.5\textwidth]{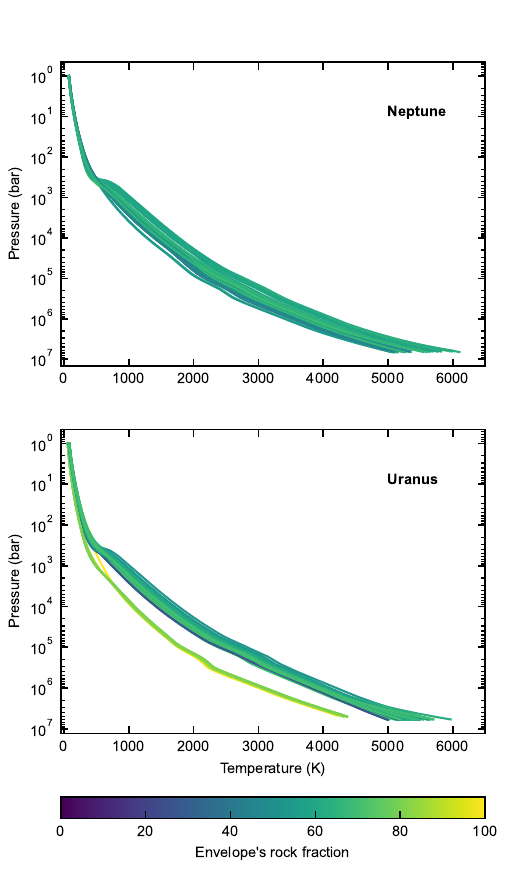}
    \caption{Impact of rock fraction on the planetary $P$–$T$ profiles. Shown are the $P$–$T$ profiles of Neptune (top) and Uranus (bottom) for a representative sample of 100 interior models that satisfy the observational constraints. The curves are color-coded by the effective global rock fraction in the envelope (relative to its total heavy element content), such that each model is represented by a single color. Note that while a single global value is shown per model, the local rock abundance is physically restricted to regions where the temperature exceeds the silicate condensation threshold and is zero at lower temperatures (as described in Sect.~\ref{sec:methods}). Models with lower envelope rock fractions, corresponding to higher ice (water) content, exhibit higher interior temperatures.}
    \label{fig:pt_profiles}
\end{figure}

Although the envelope constitutes only a minor fraction of the total planetary mass of Uranus and Neptune, making it a poor indicator of their bulk properties, its composition -- specifically the rock fraction -- has a profound influence on the thermodynamic structure of the deep interior. To quantify this effect, Fig. \ref{fig:pt_profiles} presents the pressure-temperature ($P-T$) profiles for a representative subsample of 100 posterior models for Neptune (top) and Uranus (bottom), color-coded by the rock mass fraction within the envelope's heavy component (where a high value indicates that the heavy elements in the envelope are mainly rocks rather than water).

Starting from the outer boundary, all models converge at approximately the same initial state, corresponding to the surface boundary condition at 1 bar, its associated temperature (see Table \ref{table:prop}), and the entropy that defines the adiabat. Up to pressures of several hundred bars (representing around 99\% of the planet's radius), the models follow a similar adiabat. At greater depths, significant deviations emerge. We observe an overall trend where the rock fraction in the envelope and the deep internal temperature are anticorrelated. Models with lower rock abundances (and correspondingly higher ice fractions) are intrinsically less dense and therefore require higher temperatures along the adiabat to satisfy hydrostatic equilibrium, resulting in steeper temperature gradients and hotter interiors. In contrast, envelopes enriched in rocks are denser and follow cooler temperature profiles at depth.

Temperature differences between models with varying rock fractions can exceed 1000 K in extreme cases. Such differences could directly affect electrical conductivity, phase boundaries, and the depth and efficiency of dynamo action, with potential implications for the observed magnetic field morphologies of both planets. This underscores the importance of accurately accounting for rocks when modeling the interiors of Uranus and Neptune. Many studies consider water as the only heavy component in the envelope and mantle, a simplification that can falsely suggest a hotter planetary interior than would occur if rocks were included.

A pronounced jump in the $P$–$T$ profiles occurs at approximately 0.1~Mbar for Neptune and 0.2~Mbar for Uranus, marking the transition between the envelope and the mantle. For Uranus, the solutions exhibit a structured separation that is primarily associated with the depth of this transition (Fig.~\ref{fig:pt_profiles}, bottom panel). The majority of models place the envelope–mantle boundary at pressures $P_{1-2} > 0.13$~Mbar and follow relatively hot adiabats, reaching higher temperatures in the deep interior. A smaller subset of solutions ($\sim$7\% of the accepted models) undergoes the transition at shallower pressures, $P_{1-2} < 0.13$~Mbar, and remains significantly cooler at depth.

 As discussed in Sect. \ref{sec:methods}, a transition from a low-metallicity outer envelope to a more metal-rich deeper region is necessary to match the gravitational data for Uranus and Neptune. One possible explanation for this transition is hydrogen-water immiscibility, which has been proposed to occur in the deep interior of the ice giants \citep{Bailey2021, Bergermann2024, Cano2024, Gupta2025}. However, further experimental data are needed to fully characterize this transition under the extreme conditions within planetary interiors. Our results suggest that the precise depth of this boundary is degenerate with the bulk composition of the overlying envelope. As discussed in Appendix~\ref{secA0}, scenarios where rocks are confined strictly to the core tend to force this transition to even lower pressures ($< 0.06$~Mbar for Neptune and $< 0.16$~Mbar for Uranus; see Fig.~\ref{fig:scenarios_comparison}).

\subsection{Impact of the heavy-element equations of state}
\label{sec:metals_eos}

The choice of EOSs for the heavy elements can substantially affect the inferred internal structure of Uranus and Neptune, particularly given the high metallicities expected in their interiors. To quantify this effect, we investigated the sensitivity of our inferred ice-to-rock partitioning to the adopted EOSs for both the ice (water) and rock components.

\subsubsection{Influence of the water equation of state}
\label{sec:water_eos}

\begin{figure*}
    \centering
    \includegraphics[width=\textwidth]{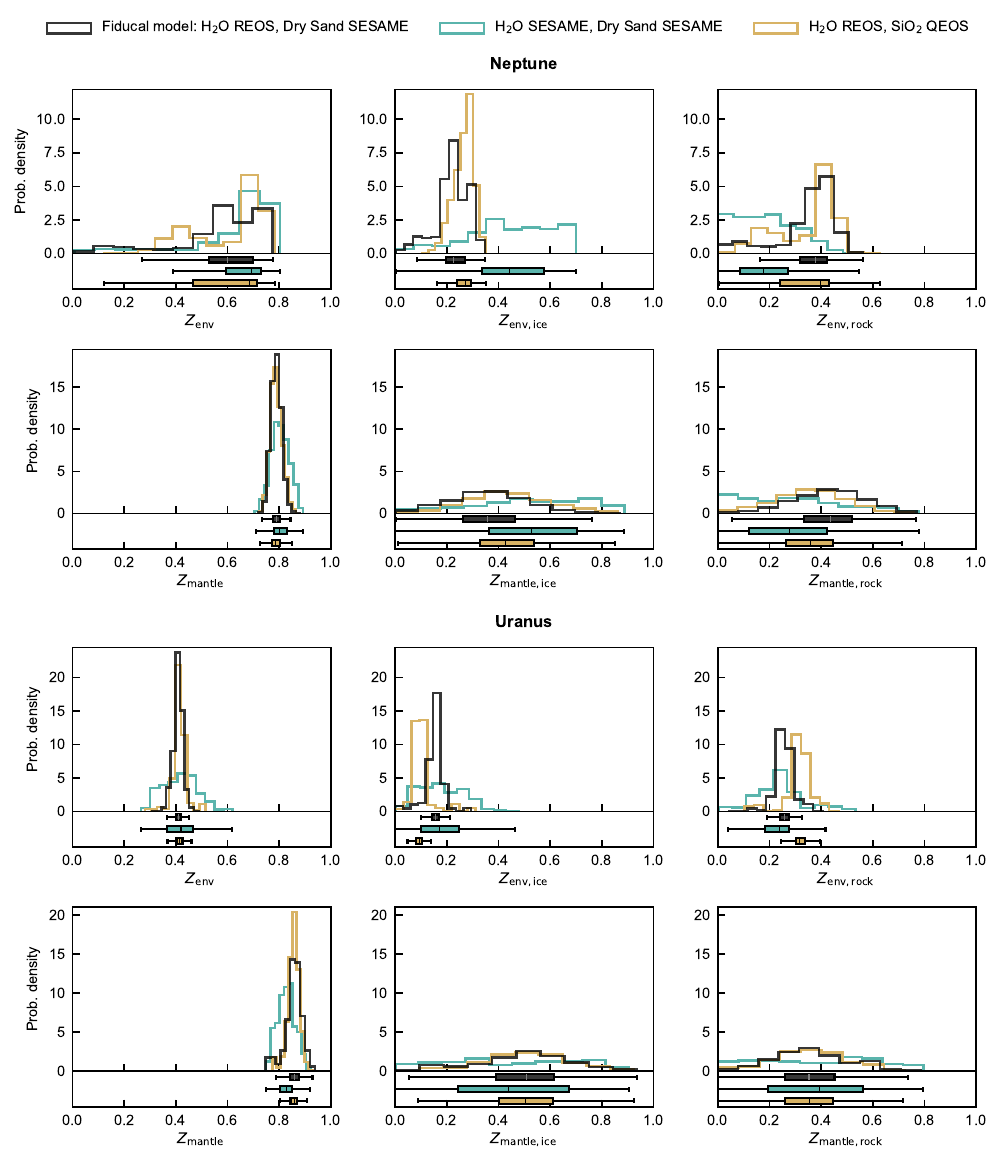}
    \caption{Impact of different EOSs for water and rock on the inferred heavy-element content. We compare the posterior distributions for Neptune (top) and Uranus (bottom), showing the envelope and mantle metallicities,as well as the corresponding ice and rock fractions. The fiducial model (black) adopts REOS for water and the SESAME Dry Sand EOS for the rocky component. This is compared against models that use an alternative water EOS (SESAME; blue) and an alternative rock EOS (QEOS for $\text{SiO}_2$; yellow). For both planets, the first row presents results for the envelope (total metallicity and the ice and rock mass mixing ratios), while the second row shows the corresponding quantities for the mantle. The lower panel of each subplot contains box plots summarizing the distributions; the central vertical line indicates the median, the box represents the IQR, and the whiskers extend to values within 1.5 times the IQR.}
    \label{fig:eos_compare}
\end{figure*}

We first performed parallel retrievals using two distinct water EOSs: REOS, calculated by \citet{Scheibe2019, Scheibe2021} based on the work of \citet{French2009} and \citet{Mazevet2019}, and the widely used SESAME EOS \citep{Lyon1992}. Figure~\ref{fig:eos_compare} displays the resulting posterior distributions for the metallicity and for the partitioning between ice and rock in the envelopes and mantles of both planets.

The two water EOSs yield comparable total metallicities for Uranus and Neptune in both the envelope and mantle; however, significant differences arise in the partitioning between ice and rock. For Neptune, the choice of EOS has a pronounced impact on the inferred envelope composition. REOS favors a rock-rich envelope, with a median rock mass mixing ratio of $Z_{\rm env,rock} \sim 0.4$ and a correspondingly low ice fraction. In contrast, the SESAME water EOS allows for substantially more ice-rich solutions, with a lower median rock fraction ($Z_{\rm env,rock} \sim 0.2$). A similar trend is found in Neptune’s mantle: models employing REOS preferentially peak at higher rock fractions, whereas the SESAME water EOS permits more ice-dominated compositions. For Uranus, the differences between the two EOSs are more modest in both the envelope and mantle, although REOS provides tighter constraints on the envelope composition.

To better understand the physical origin of these differences, Fig.~\ref{fig:density_profiles} compares the density–temperature profiles for Uranus and Neptune. For each planet, the interior profiles are computed using the same set of structural parameters (core mass, metallicities, transition pressure, and ice-to-rock ratios) drawn from a representative posterior solution, ensuring that the profiles reflect viable interior conditions and that any differences in density arise solely from the adopted EOS. For both planets, the SESAME water EOS systematically predicts higher densities than REOS (dashed blue lines vs. solid black lines) over the temperature range relevant to the deep envelope and mantle. This behavior directly explains the shifts in the inferred ice-to-rock partitioning. Because the SESAME water EOS is intrinsically denser under these conditions, interior models can satisfy the gravitational constraints with a larger fraction of ice and a reduced contribution from rock. In contrast, the lower densities predicted by REOS require a higher mean heavy-element density to reproduce the same observational data, which is achieved by favoring more rock-rich compositions. This interpretation is consistent with the findings of \citet{Scheibe2019}, who showed that interior models using REOS require a higher heavy-element content to match the planet’s radius.

\subsubsection{Influence of the rock equation of state}
\label{sec:rock_eos}

Given our finding that both planets' envelopes are systematically rock-enriched, the assumed rock EOS is also expected to have an effect on the inferred composition. To test this, we compared our fiducial models, which utilize the SESAME Dry Sand EOS for the rocky component, with models employing QEOS for SiO$_2$ \citep{More1988, Vazan2013}.

Similar to our water EOS comparison, the retrieved total metallicities remain broadly consistent across both the envelope and mantle regardless of the chosen rock EOS (Fig.~\ref{fig:eos_compare}). While the exact partitioning between ice and rock shows some sensitivity to this choice, the overall shifts are relatively modest. For Neptune, both the fiducial (SESAME Dry Sand) and alternative (QEOS for SiO$_2$) models consistently favor rock-enriched envelopes. In the mantle, however, the SESAME EOS yields slightly higher rock fractions than QEOS. In Uranus, the inferred mantle composition remains highly robust between the two models, whereas QEOS drives the envelope toward a slightly rockier composition compared to our fiducial setup.

\begin{figure}[t!]
    \vspace{-25pt}
    \centering
            \includegraphics[width=0.5\textwidth]{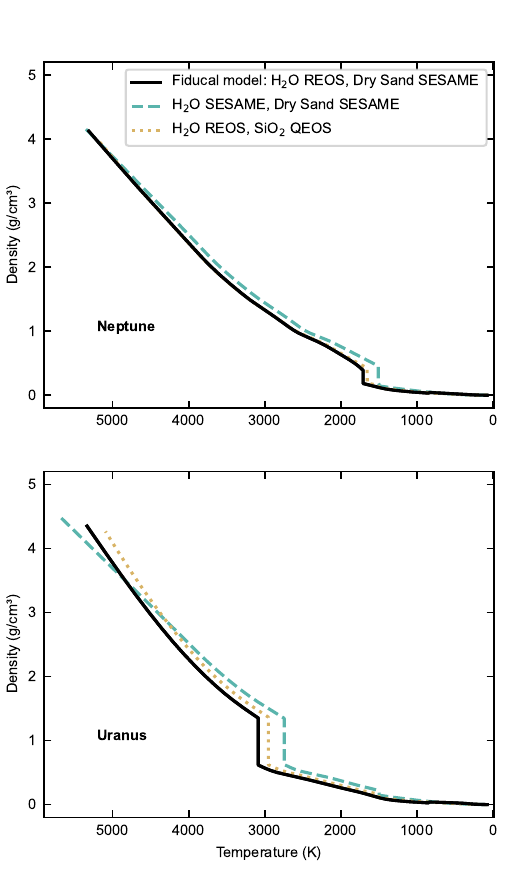}
    \caption{Density as a function of temperature for Neptune (top) and Uranus (bottom). The profiles were calculated using our fiducial model (REOS for water and SESAME Dry Sand for rock; solid black lines) and two alternative test models: one substituting the water EOS with SESAME (dashed blue lines) and another substituting the rock EOS with QEOS for $\text{SiO}_2$ (dotted yellow lines). To isolate the effect of the EOS, the profiles for each planet were computed using identical interior parameters drawn from a representative posterior solution. The SESAME water EOS yields denser profiles than the fiducial REOS throughout most of the interior. Similarly, using QEOS for $\text{SiO}_2$ instead of SESAME Dry Sand results in denser structural profiles, with the differences being most prominent in Uranus.}
    \label{fig:density_profiles}
\end{figure}

The rock EOS introduces a comparable structural adjustment, driven by the thermodynamic properties of the chosen tables. As shown in Fig.~\ref{fig:density_profiles}, replacing the SESAME Dry Sand EOS with QEOS for SiO$_2$ (dotted yellow lines) modifies the deep-interior density profile, particularly for Uranus. Although QEOS is intrinsically less dense than the SESAME Dry Sand EOS at similar thermodynamic conditions, the requirement to reproduce the observed mass and radius leads the structure solver to converge toward a slightly colder adiabat, resulting in a denser self-consistent interior profile. In the full MCMC retrieval, where the composition is allowed to vary freely, this structural requirement is instead accommodated by increasing the rock fraction and decreasing the ice fraction in the envelope of Uranus to accurately reproduce the observational constraints.

Importantly, while the exact inferred ice-to-rock partitioning depends on the adopted heavy-element EOS, all tested EOS combinations consistently indicate that both Uranus and Neptune possess envelopes that are substantially enriched in heavy elements relative to pure-ice compositions (see Appendix~\ref{secA0}).

\section{Discussion}\label{sec:discussion}

\subsection{Uranus and Neptune discrepancies}

Our results indicate fundamental differences in the internal architectures of Uranus and Neptune, challenging the traditional view of these planets as compositional twins. While both planets exhibit highly enriched mantles ($Z_{\rm mantle} > 0.7$), their envelopes differ markedly. Uranus is generally reproduced by models with a moderately enriched envelope (median $Z_{\rm env} \sim 0.4$) with a correspondingly larger hydrogen mass fraction, whereas Neptune favors substantially more metal-rich envelopes (median $Z_{\rm env} \sim 0.6$). These trends persist across our explored parameter space and EOS choices, despite the inherent degeneracies of planetary interior modeling.

A key outcome of our analysis is the substantial refractory content for the outer layers of both planets. Within the class of three-layer models considered here, the envelopes are consistently enriched in rocks relative to ices, with median relative rock fractions exceeding 60\% of the heavy-element component. This challenges the traditional interpretation of Uranus and Neptune as predominantly ice-dominated planets and instead implies ice-to-rock ratios comparable to those inferred for other outer Solar System bodies, such as Pluto, Kuiper belt objects, and comets, where rock fractions can approach $\sim$70\%.

The compositional dichotomy between the two planets extends to the distribution of heavy elements. Neptune exhibits higher total mass mixing ratios of rock in its envelope, whereas Uranus retains a more substantial hydrogen reservoir. Although mantle compositions are less tightly constrained, a similar divergence is apparent at depth: Uranus solutions tend toward higher overall metallicities with a larger ice fraction, whereas Neptune’s mantle is preferentially rock-rich. While these results remain subject to degeneracies in interior structure modeling, they consistently point to distinct internal architectures for the two planets. These contrasting compositions underscore the unique and complex internal structures of these planets, and raise important questions about whether Uranus and Neptune experienced distinct formation and evolutionary pathways that have led to the differences observed today.

The observed disparity in rock content between the two planets may reflect differences in their accretion histories. The elevated rock abundance in Neptune’s envelope and mantle could indicate more efficient accretion of small bodies, such as planetesimals, asteroids, or comets. This rocky material could remain segregated from the deeper interior in the absence of effective mixing mechanisms--a phenomenon recently proposed for Jupiter \citep{Howard2023, Muller2024}. Such a process could naturally give rise to the vertical gradients in rock fraction inferred in our models, particularly for Uranus.

Giant impacts provide another plausible mechanism for producing the heavy-element enrichment inferred here. Collisions with larger objects, such as planetary embryos, have been invoked to explain the high metallicities of the ice giants, as well as their distinct axial tilts, satellite systems, and inferred moments of inertia \citep{Podolak2012, Reinhardt2020, Chau2021, Valletta2022}. A sufficiently energetic impact could deposit a substantial mass of refractory material directly into the envelope, potentially accounting for the high rock fractions we derived. The contrasting density profiles and heat fluxes of Uranus and Neptune \citep{Pearl1990, Pearl1991} further point to different degrees of post-formation mixing, which may help explain the more uniform heavy-element distribution inferred for Neptune relative to the more stratified solutions favored for Uranus. These interpretations remain speculative but are qualitatively consistent with our inferred compositional trends.

Beyond formation, the compositional differences between Uranus and Neptune may also reflect ongoing evolutionary processes, particularly phase separation among constituent materials such as H$_2$–H$_2$O, which remains poorly understood. \citet{Bailey2021} proposed that the differences in envelope metallicity could be explained if Neptune is currently at an earlier stage of H$_2$-H$_2$O de-mixing than Uranus, based on models in which rocks are confined to the core. Our results expand on this hypothesis by incorporating rocks to the envelope and mantle. We suggest that Neptune’s high envelope metallicity is driven not only by its earlier de-mixing stage but also by a substantial rock component. Conversely, the lower water abundance in the envelope of Uranus relative to its mantle may indicate a more advanced stage of hydrogen–water separation, where a larger fraction of water has rained out into the deep interior. De-mixing processes play a significant role on other planets as well; for instance, on Jupiter and Saturn, helium separation from metallic hydrogen under high pressures results in “helium rain” \citep{Mankovich2020, Howard2024}, which may explain atmospheric depletion of helium and neon \citep{vonZahn1998, Militzer2010}. Similar de-mixing processes could influence the internal distributions of elements in Uranus and Neptune, although the exact mechanisms and stages may vary. Further investigation of these processes would provide valuable insights into the interiors of these planets.

Constraining ice and rock fractions in Uranus and Neptune is therefore critical for reconstructing their formation and evolution histories. The precise location at which these planets formed -- whether closer to the Sun, between 5 and 15 AU \citep{Levison2015}, or farther out, between 12 and 30 AU \citep{Helled2014} -- remains uncertain. Their formation location would significantly affect their accretion rates and resulting compositions \citep{Izidoro2015}, making robust constraints on interior structure essential for distinguishing among competing formation scenarios. Resolving these structural uncertainties is vital not only for understanding the origins of the Solar System's ice giants, but also for comparative studies with similar exoplanets.

\subsection{Model inputs and limitations}

While our MCMC framework extensively explores the parameter space of planetary interiors, our inferences are inherently tied to our underlying structural assumptions and model inputs. A critical assumption in our three-layer models is the presence of an envelope-mantle structure, which is physically motivated by the immiscibility of hydrogen and water at high pressures and temperatures \citep[e.g.,][]{Bailey2021, Bergermann2024, Cano2024, Gupta2025}. The pressure at which this transition takes place is a model parameter ($P_{1-2}$) that is explored in our Bayesian framework. Because this boundary separates two compositionally distinct regions, its location directly influences the distribution of heavy elements and the resulting ice and rock fractions. Specifically, $P_{1-2}$ dictates the relative masses of the envelope and the mantle, which naturally limits the total inventory of heavy elements each layer can accommodate. Furthermore, the transition pressure establishes the thermal and pressure conditions of the envelope; if $P_{1-2}$ is located at pressures lower than the silicate cloud base, rocky material is physically restricted from existing in the envelope and must be entirely sequestered in the deeper interior. These interdependencies are reflected in our posterior distributions: as shown in our MCMC corner plots (see Appendix B, e.g., Figs.~\ref{fig:corner_neptune_rocks_reos} and \ref{fig:corner_uranus_rocks_reos}), the inferred heavy-element mass fractions in both the envelope and the mantle exhibit clear correlations with the transition pressure $P_{1-2}$.

Closely related to the assumption of a distinct envelope-mantle transition is our parametrization of the deepest interior. Our models assume a classical structure consisting of an outer envelope, an inner mantle, and a distinct rocky core. While this parametrization provides a convenient framework for exploring the allowed compositions, it is not unique. Interior models of Jupiter and Saturn strongly suggest the presence of dilute cores \citep{Wahl2017, Mankovich2021, Miguel2022}, where heavy elements gradually mix with hydrogen and helium deep within the planet rather than residing beneath a sharp compositional boundary. If such a structure is present in Uranus or Neptune, some fraction of the rocky material currently assigned to a compact core in our models would instead be distributed throughout the deep interior \citep[e.g.,][]{Neuenschwander2024, Malamud2024, Morf2025}. This topology would naturally alter the distribution of heavy elements and the resulting ice-to-rock ratios, depending on the radial extent of the dilute region and the efficiency of mixing. However, while a dilute core is a physically plausible scenario for the ice giants, the available data for Uranus and Neptune are far more limited than for Jupiter and Saturn, where high-precision gravity harmonics and, in Saturn’s case, ring seismology provide strong constraints on internal structure. As a result, current data are insufficient to robustly constrain the presence or extent of a diluted core in the ice giants. Exploring dilute core structures remains beyond the scope of this work, but it represents an important avenue for future studies aimed at understanding the internal structure and formation histories of Uranus and Neptune.

Beyond these internal structural boundaries, our interpretation of the gravitational constraints is influenced by assumptions regarding atmospheric dynamics. The measured gravitational moments ($J_2$ and $J_4$) of Uranus and Neptune represent a combination of the static planetary figure in hydrostatic equilibrium and dynamic contributions from atmospheric zonal winds. Because our models assume a purely hydrostatic interior, we neglected this wind-induced component. Despite the limited data available, previous studies have constrained the inward extent of these winds and their subsequent effect on the gravity field. For instance, \citet{Kaspi2013} inferred that atmospheric dynamics are restricted to the outermost $0.15$\% and $0.2$\% of the total mass for Uranus and Neptune, respectively, establishing an upper limit of $\sim$1000~km for the depth of the dynamical atmosphere. Similarly, \citet{Soyuer2023} found that the maximum decay scale heights of these zonal winds correspond to $2-3$\% of the planetary radii. The wind-induced contributions to both $J_2$ and $J_4$ are estimated to be on the order of $10^{-6}$ or less \citep{Kaspi2013, Soyuer2023}. For context, the current $1\sigma$ observational errors for $J_2$ and $J_4$ are of a similar order of magnitude (Table~\ref{table:prop}). Thus, the missing dynamical signals are generally comparable to or smaller than the current measurement uncertainties. For comparison, recent studies indicate that while Jupiter's zonal winds are confined to the outer $3$--$4$\% of the planetary radius, they still produce a measurable effect on its derived heavy-element distribution \citep{Wahl2017, Kaspi2018, Debras2019}. A direct extrapolation to Uranus and Neptune, however, is not straightforward. Their wind profiles differ significantly from those of the gas giants, particularly in the equatorial region \citep[e.g.,][]{Sromovsky2015, Tollefson2018, Helled2020}. Furthermore, given the substantially larger inherent uncertainties in modeling their interiors, any systematic bias introduced by neglecting differential rotation is likely overshadowed by broad structural degeneracies to a much greater degree than for Jupiter. Consequently, the static hydrostatic assumption remains a robust approximation for our present interior inferences.

\section{Conclusions}\label{sec:conclusions}

We have conducted a comprehensive Bayesian retrieval of the internal structures of Uranus and Neptune, utilizing three-layer models to explore the full parameter space of their compositions. Our analysis challenges the conventional ``ice-dominated'' paradigm, indicating that the envelopes of both planets likely harbor significant reservoirs of refractory materials. The inferred rock mass fractions in the outer layers align closely with the composition of outer Solar System objects.

Crucially, our results delineate a clear architectural bifurcation between the two planets. Uranus is best described by a more differentiated structure and has retained a higher hydrogen mass fraction in a moderately enriched envelope on top of a distinct, high-metallicity mantle. In contrast, Neptune has a more thoroughly mixed profile, characterized by a highly metallic envelope and a deep interior where rocks may dominate over ices. These compositional disparities support the hypothesis that despite their similar masses and radii, Uranus and Neptune experienced distinct formation and evolutionary pathways, potentially involving different accretion histories or distinct regimes of post-formation phase separation.

Methodologically, we have demonstrated that allowing for rocks within the envelope and mantle -- rather than confining them to the core -- drastically alters the retrieved metallicities, $P-T$ profiles, and the depth of compositional transitions. Furthermore, we find that the choice of the water EOS introduces non-negligible systematic uncertainties, highlighting the urgent need for accurate experimental data at relevant thermodynamic conditions. Altogether, our study highlights the importance of moving beyond simplified ice-dominated models and the need for in situ measurements to resolve current degeneracies in the interior structure of these planets.

Future work should further explore how both rock-rich and ice-rich scenarios impact planetary formation models; this would provide insights into the Solar System’s history and help identify potential differences in the formation and evolution of Uranus and Neptune. Although the two planets have comparable physical characteristics, their unique compositions emphasize the need for dedicated studies. A targeted mission to the ice giants would be invaluable \citep{Fletcher2022, Mousis2022, Movshovitz2022, Girija2023, Mandt2023, Mousis2024, Morf2024, Lin2025}, not only to refine our understanding of their interiors but also to allow us to place Uranus and Neptune within the broader context of the diverse population of intermediate-mass exoplanets.

\section*{Data availability}
The posterior samples from our interior models of Uranus and Neptune are publicly available on GitHub at \url{https://github.com/AstroYamila-Team/UranusNeptune_Interiors}. The CEPAM code is freely available at \url{https://svn.oca.eu/codes/CEPAM}.

\begin{acknowledgements}
The authors thank the anonymous referee for their constructive comments that helped to improve the manuscript. This project has received funding from the European Research Council (ERC) under the European Union’s Horizon 2020 research and innovation programme (grant agreement no. 101088557, N-GINE). This work used the Dutch national e-infrastructure with the support of the SURF Cooperative using grant no. EINF-15103. The authors thank Nadine Nettelmann and Ludwig Scheibe for providing the water equation of state used in their work, and Simon M\"uller for providing the QEOS table for SiO$_2$.
\end{acknowledgements}

\bibliographystyle{aa}
\bibliography{ref}

\begin{appendix}

\section{Test scenario: Confining rocks to the core}\label{secA0}

\begin{figure*}
    \centering
    \includegraphics[width=\textwidth]{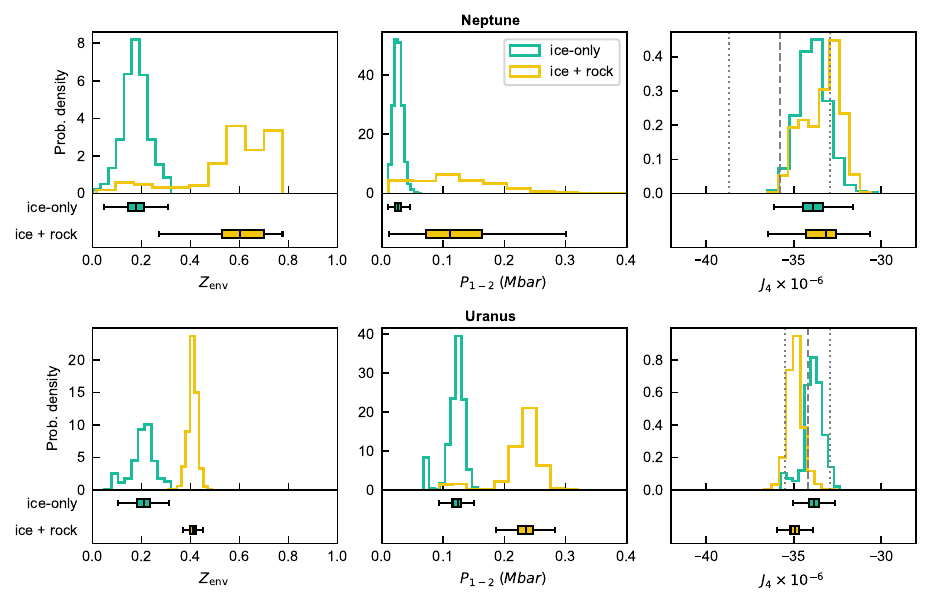}
    \caption{Comparison of key parameters to differentiate between the ice-only and ice+rock scenarios for Neptune and Uranus. Left: Envelope metallicity. 
    Middle: Pressure at the transition between the envelope and mantle. 
    Right: $J_4$ gravitational harmonic, including the corresponding measured value (dashed line) and its 1$\sigma$ error range (dotted lines). Bottom subpanels: Box plots summarizing the distributions, where the central vertical line indicates the median, the box represents the IQR, and the whiskers extend to values within 1.5 times the IQR.}
    \label{fig:scenarios_comparison}
\end{figure*}

We investigated a hypothetical scenario in which rocks are confined to the planetary core, such that ices (assumed to be pure water) are the only heavy elements present in the envelopes and mantles of Uranus and Neptune. Comparing these ice-only models with our fiducial ice+rock models (Fig.~\ref{fig:scenarios_comparison}) reveals that neglecting rocks systematically leads to lower envelope metallicities and shifts the envelope--mantle transition to lower pressures. This demonstrates that the inclusion of rocky material has a substantial impact on the inferred internal structures of both planets.

A particularly diagnostic difference between the two scenarios is the depth of the envelope--mantle transition in both planets. For Neptune, the ice-only models predict very shallow transitions: in 99.6\% of the accepted solutions, the transition occurs at pressures below 0.05~Mbar. In contrast, in the ice+rock scenario, only 17\% of the solutions place the transition below this value. Such shallow transitions appear inconsistent with the results of \citet{Cano2024}, who constrain the interface between water-poor and water-rich layers in Neptune to occur at pressures of at least 5~GPa (0.05~Mbar), based on H$_2$--H$_2$O de-mixing inferred from experimental and computational data. For Uranus, by contrast, the transition pressures in both the ice-only and ice+rock models are broadly consistent with the lower limit of $\gtrsim$10~GPa proposed by \citet{Cano2024}, with only 8\% and 0.8\% of the accepted solutions falling below this threshold, respectively.

\begin{figure*}
\centering 
\includegraphics[width=\textwidth]{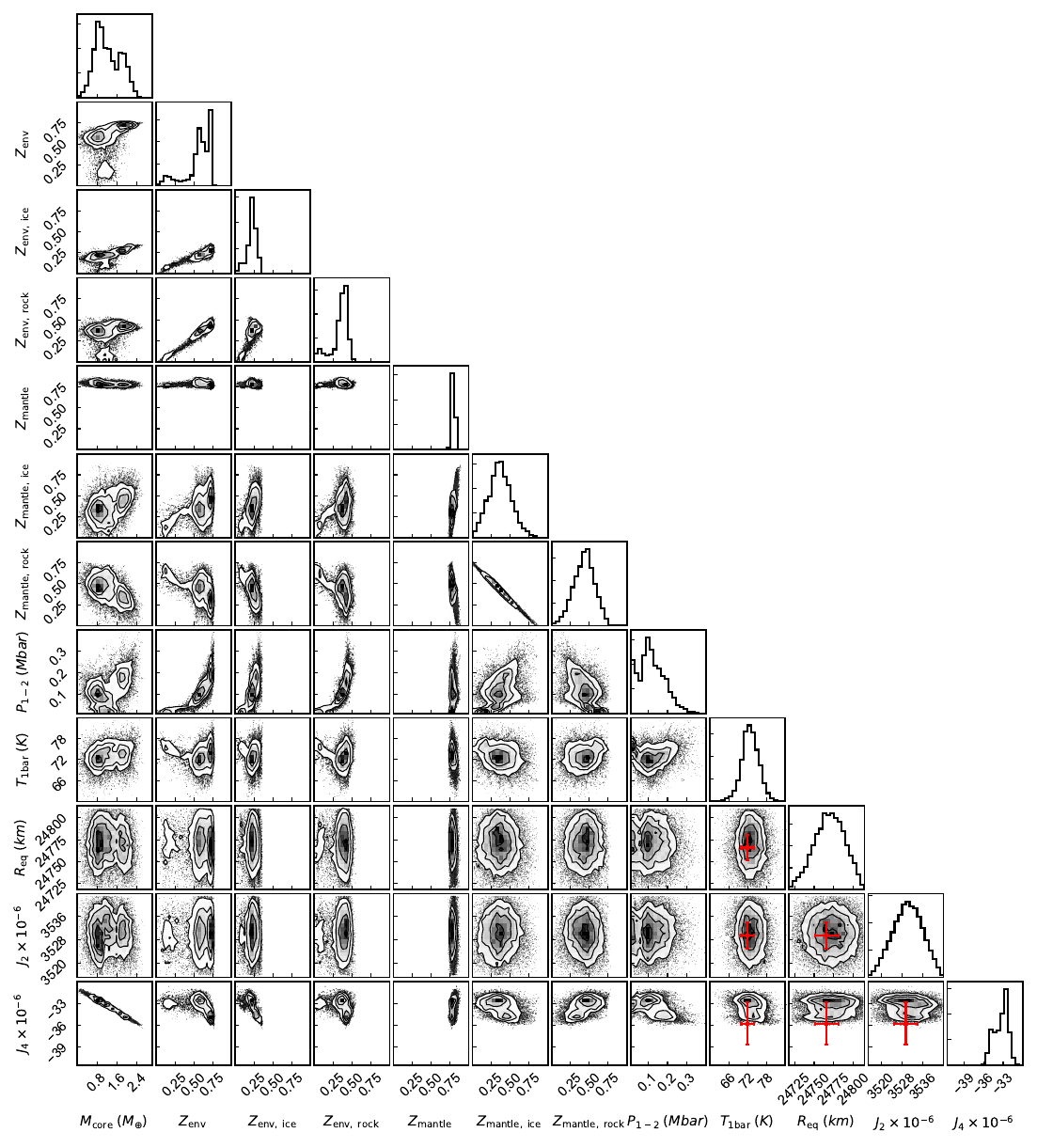}
\caption{MCMC corner plots for the fiducial ice+rock model of Neptune, adopting REOS for water and the SESAME Dry Sand EOS for the rocky component. These plots illustrate the posterior distributions of key parameters, providing insights into the relationships between different model variables and their uncertainties. Red points with error bars show the observed parameters as a reference.}
\label{fig:corner_neptune_rocks_reos}
\end{figure*}

\begin{figure*}
\centering 
\includegraphics[width=\textwidth]{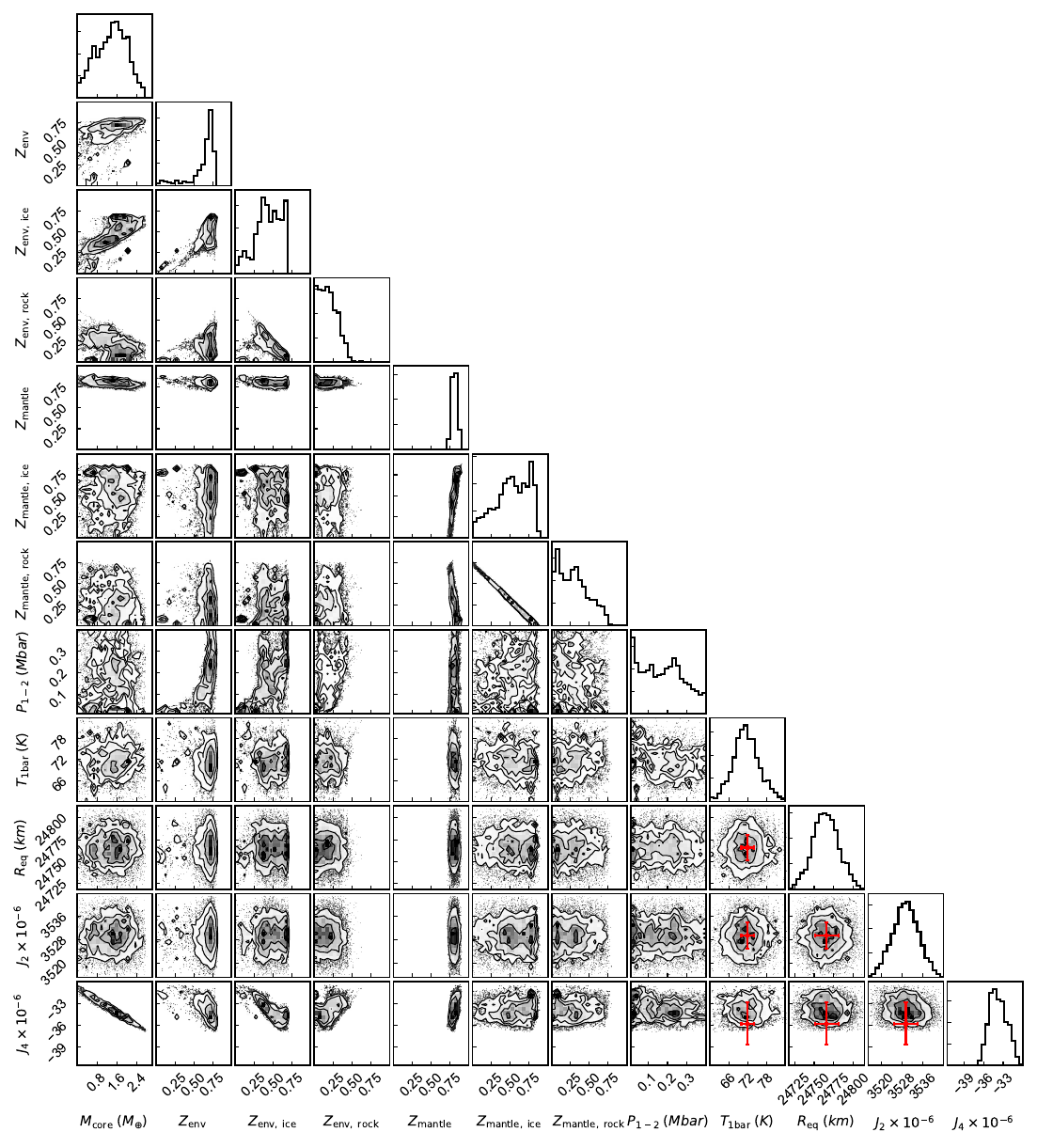}
\caption{Same as Fig.~\ref{fig:corner_neptune_rocks_reos} but for the alternative ice+rock model of Neptune, adopting the SESAME water EOS and the SESAME Dry Sand EOS for the rocky component.}
\label{fig:corner_neptune_rocks_ses}
\end{figure*}

\begin{figure*}
\centering 
\includegraphics[width=\textwidth]{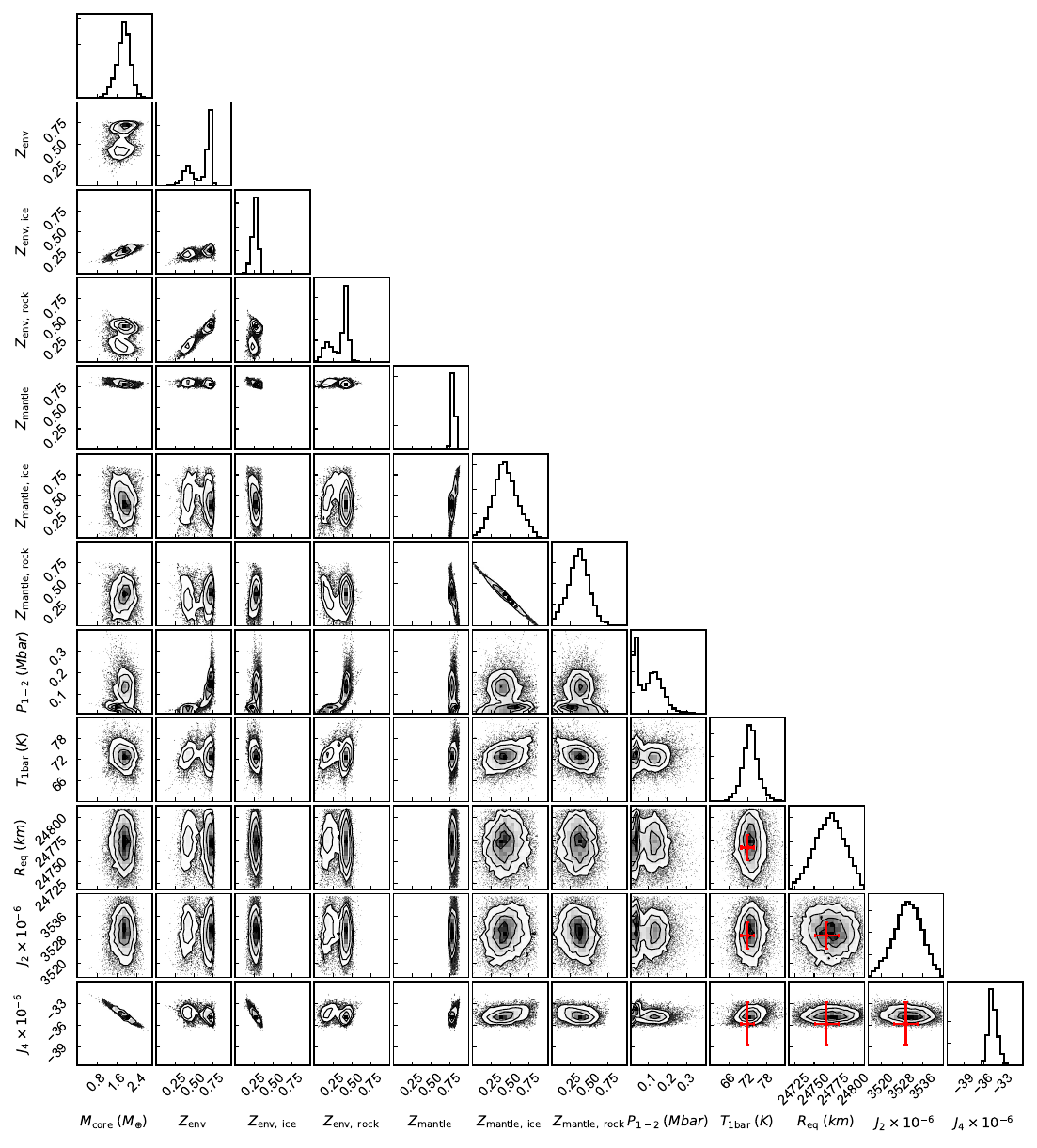}
\caption{Same as Fig.~\ref{fig:corner_neptune_rocks_reos} but for the alternative ice+rock model of Neptune, adopting REOS for water and QEOS for SiO$_2$ for the rocky component.}
\label{fig:corner_neptune_rocks_qeos}
\end{figure*}

\begin{figure*}
\centering 
\includegraphics[width=\textwidth]{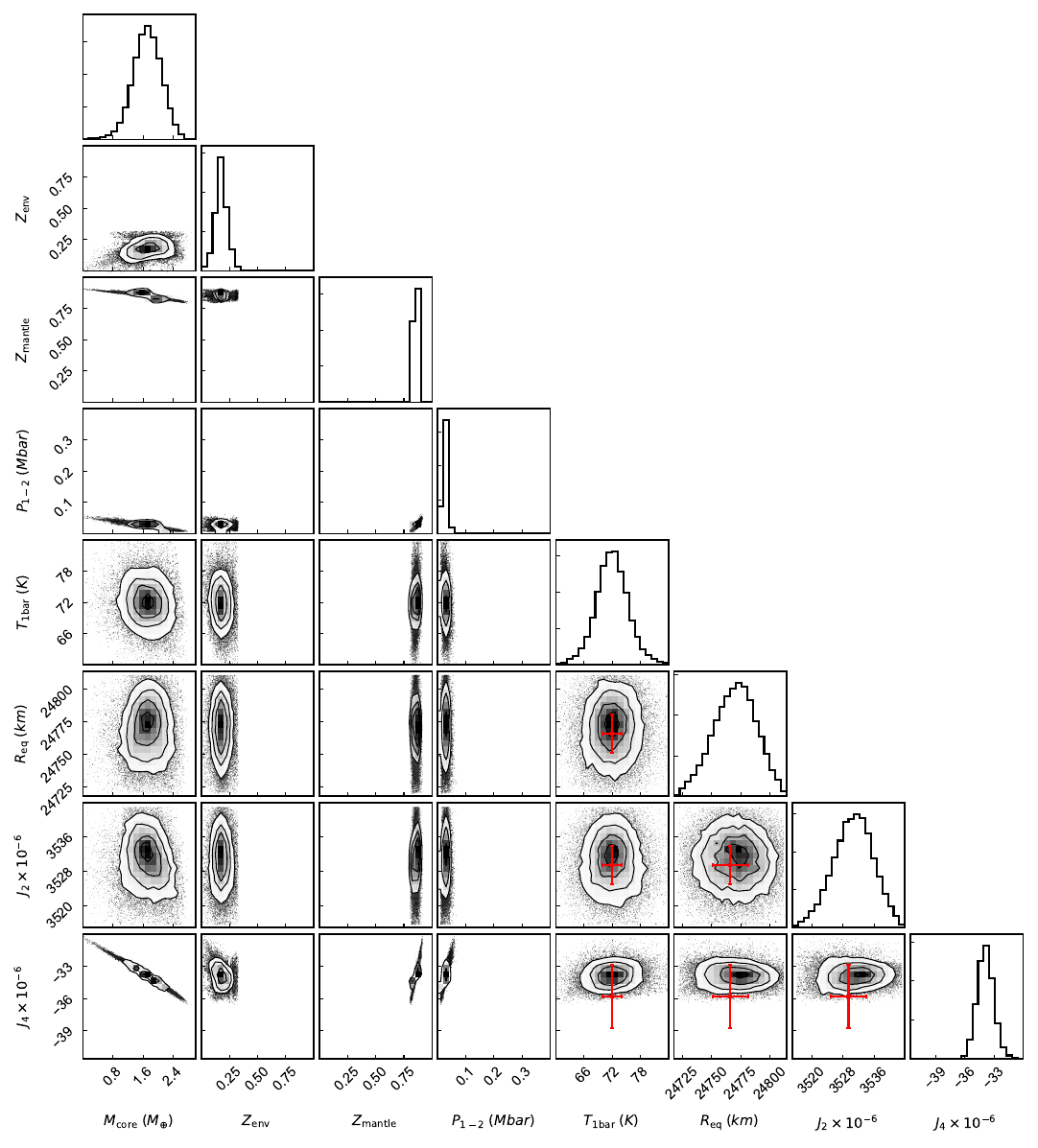}
\caption{Same as Fig.~\ref{fig:corner_neptune_rocks_reos} but for the ice-only test model of Neptune, adopting REOS for water.}
\label{fig:corner_neptune_iceonly_reos}
\end{figure*}

\begin{figure*}
\centering 
\includegraphics[width=\textwidth]{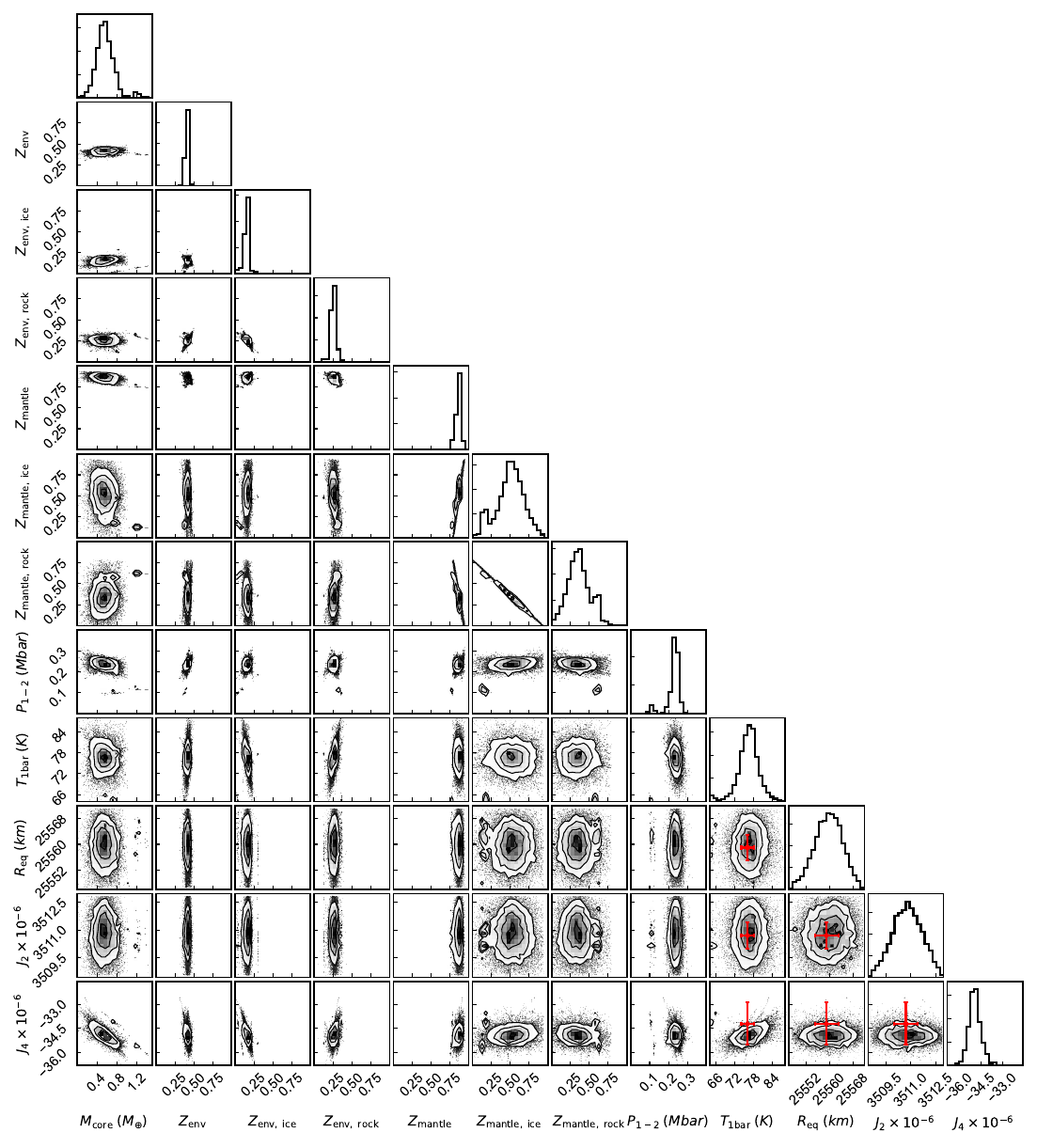}
\caption{Same as Fig.~\ref{fig:corner_neptune_rocks_reos} but for the fiducial ice+rock model of Uranus, adopting REOS for water and the SESAME Dry Sand EOS for the rocky component.}
\label{fig:corner_uranus_rocks_reos}
\end{figure*}

\begin{figure*}
\centering 
\includegraphics[width=\textwidth]{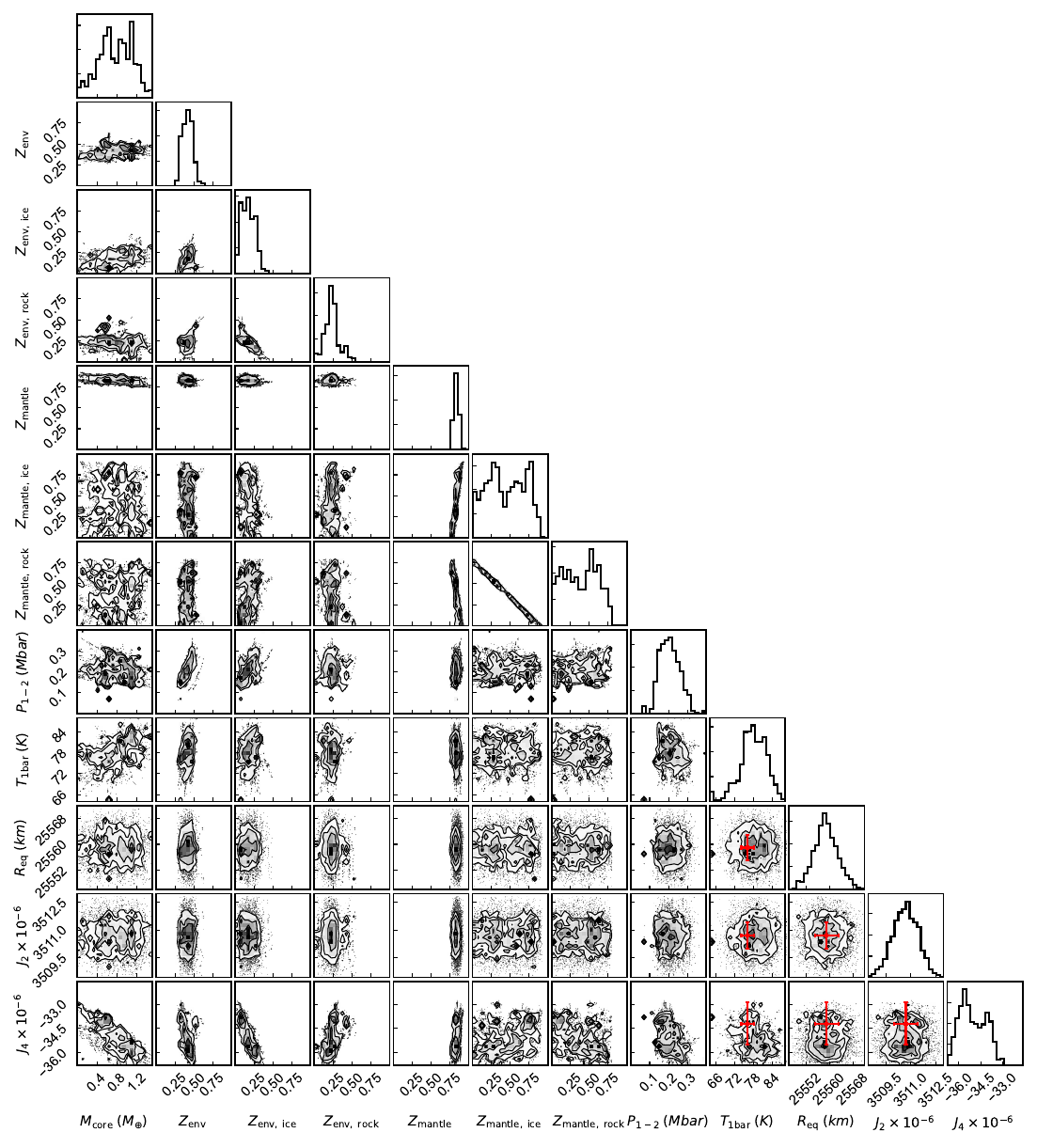}
\caption{Same as Fig.~\ref{fig:corner_neptune_rocks_reos} but for the alternative ice+rock model of Uranus, adopting the SESAME water EOS and the SESAME Dry Sand EOS for the rocky component.}
\label{fig:corner_uranus_rocks_ses}
\end{figure*}

\begin{figure*}
\centering 
\includegraphics[width=\textwidth]{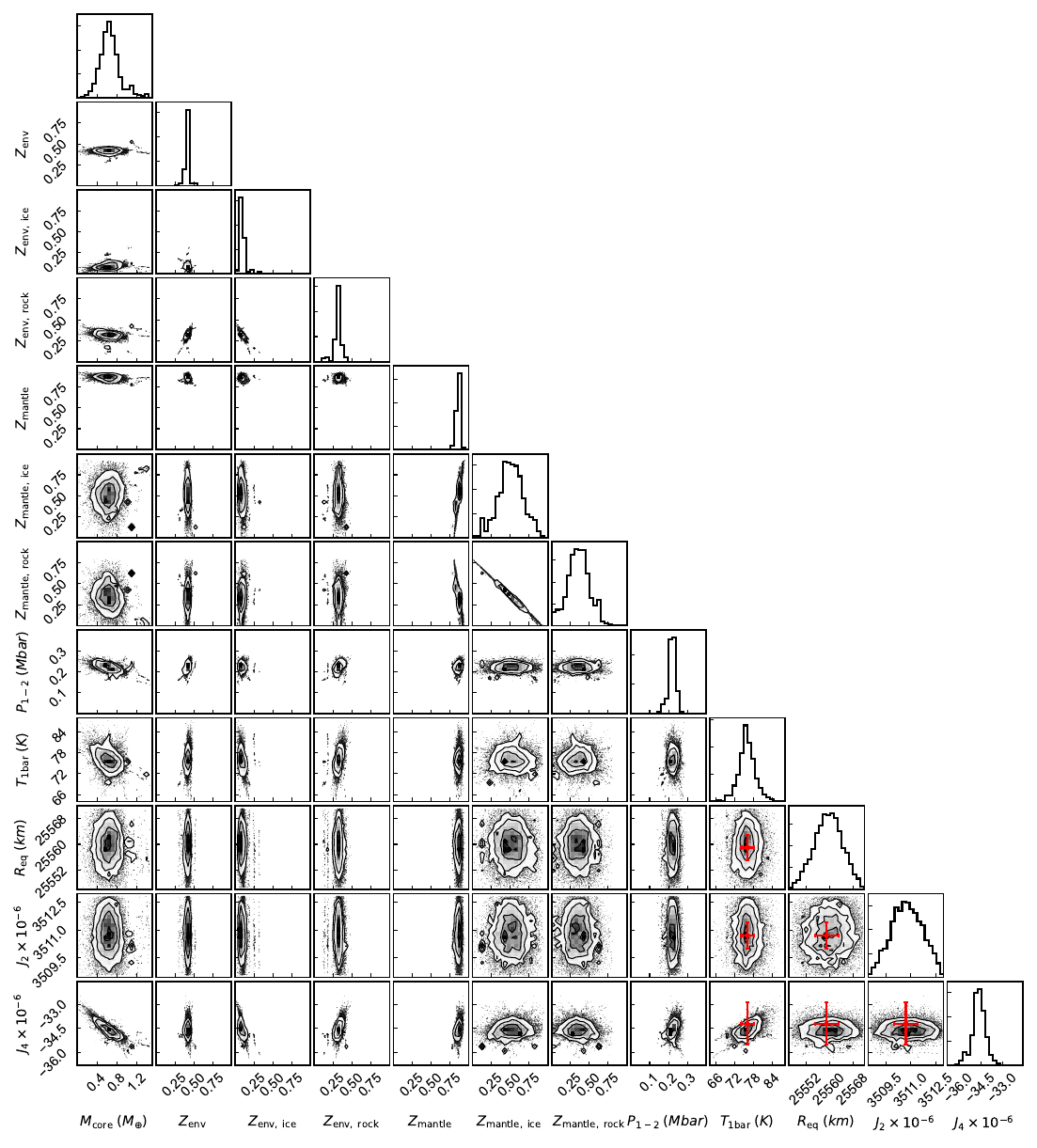}
\caption{Same as Fig.~\ref{fig:corner_neptune_rocks_reos} but for the alternative ice+rock model of Uranus, adopting REOS for water and QEOS for SiO$_2$ for the rocky component.}
\label{fig:corner_uranus_rocks_qeos}
\end{figure*}

\begin{figure*}
\centering 
\includegraphics[width=\textwidth]{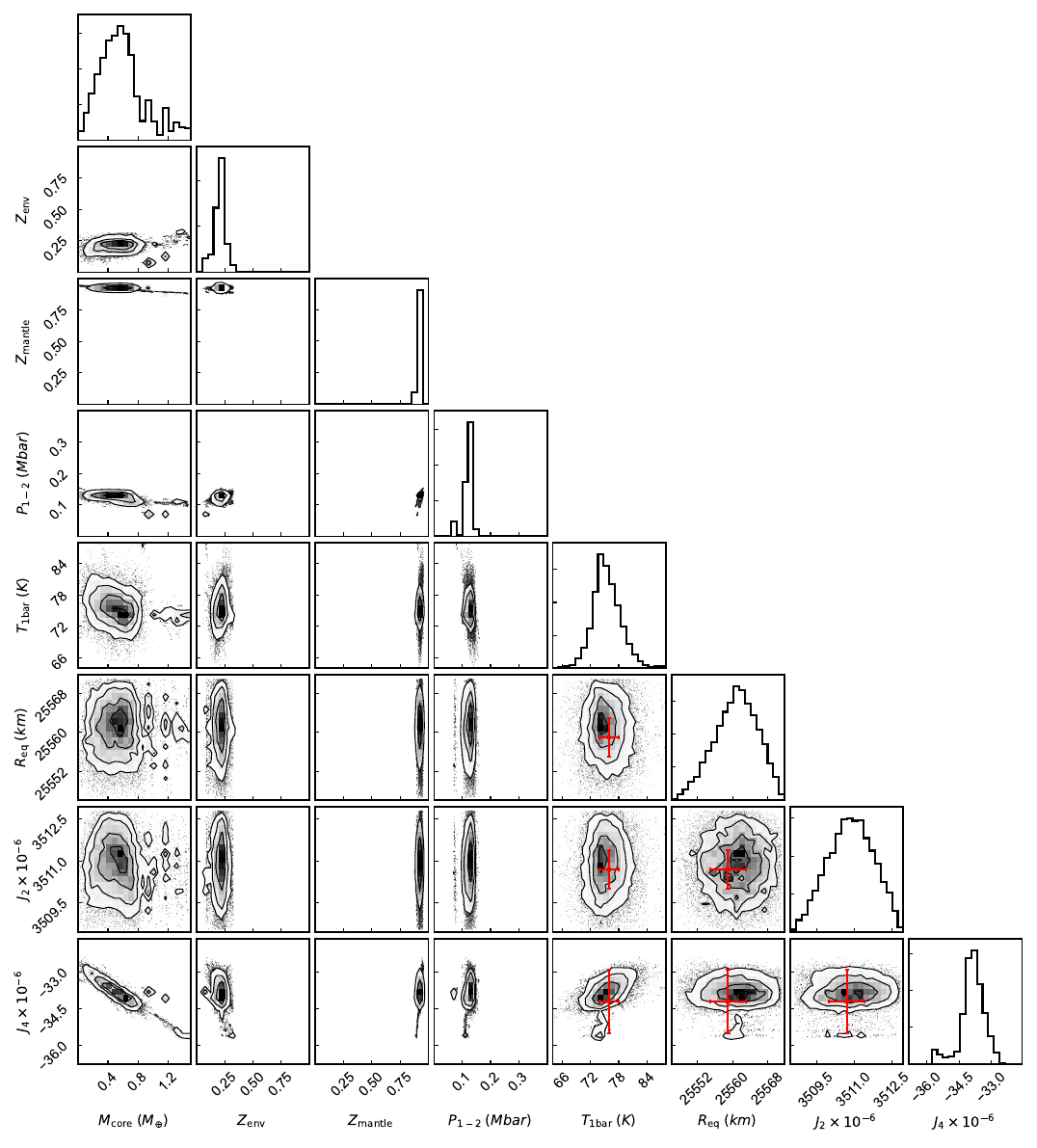}
\caption{Same as Fig.~\ref{fig:corner_neptune_rocks_reos} but for the ice-only test model of Uranus, adopting REOS for water.}
\label{fig:corner_uranus_iceonly_reos}
\end{figure*}

Both scenarios are able to reproduce the available observational constraints (see the corner plots in Figs.~\ref{fig:corner_neptune_rocks_reos}--\ref{fig:corner_uranus_iceonly_reos}), but systematic differences emerge in the fourth-order gravitational harmonic, $J_4$, particularly for Uranus. The ice-only solutions preferentially cluster near the +1$\sigma$ region, whereas the ice+rock solutions tend to lie closer to the $-1\sigma$ region. This offset complicates a definitive discrimination between the two scenarios, and improved constraints on $J_4$ alone are unlikely to fully resolve this degeneracy. For Neptune, both scenarios instead cluster near the +1$\sigma$ region of $J_4$, making this parameter less effective at distinguishing between them.

Further insights may come from more precise atmospheric abundance measurements, especially in light of the distinct envelope metallicities predicted by the two scenarios (Fig.~\ref{fig:scenarios_comparison}). Higher heavy-element abundances in the envelope would favor the ice+rock scenario, implying that both ices and rocks are present throughout the interiors of Uranus and Neptune. Conversely, lower envelope metallicities would support water as the primary heavy element, consistent with the ice-only case. Additional constraints could also be provided by measurements of higher-order gravitational moments such as $J_6$, which would help in distinguishing between these scenarios. Moreover, determining the rock fractions in the interiors of the ice giants could be instrumental in evaluating these scenarios, with approaches such as the one proposed by \citet{Nimmo2024} offering a pathway to achieve this. Finally, a better understanding of hydrogen--water immiscibility, and the potential role of other immiscibility processes, may yield further insight into the physical plausibility of each scenario.

Our ice-only results are broadly consistent with those of \citet{Nettelmann2013}, who similarly confined rocks to the core. For Uranus, \citet{Nettelmann2013} reported tightly constrained solutions characterized by low envelope metallicity ($Z_{\rm env} < 0.08$), high mantle metallicity ($Z_{\rm mantle} > 0.9$), and a small core mass ($M_{\rm core} < 0.1\,M_{\oplus}$). These findings are qualitatively consistent with our results (Fig.~\ref{fig:corner_uranus_iceonly_reos}), although we infer systematically higher envelope metallicities, reaching up to $Z_{\rm env} \sim 0.32$. For Neptune, \citet{Nettelmann2013} obtained less tightly constrained solutions, with mantle metallicities exceeding 0.65 and core masses ranging from zero to $\sim$3~$M_{\oplus}$, in agreement with our results. However, their solutions have a broader range of envelope metallicities (0--0.65), whereas in our ice-only scenario $Z_{\rm env}$ remains consistently below $\sim$0.32. This discrepancy likely arises from two main factors. First, the EOSs differ between the studies, which, as discussed in Sect.~\ref{sec:metals_eos}, can have a significant impact on the inferred internal structure. Second, the exploration of the transition pressure $P_{1-2}$ is distinct: while \citet{Nettelmann2013} allowed $P_{1-2}$ to extend to significantly greater depths (up to $\sim$1.5~Mbar), we restricted it to pressures below 0.4~Mbar, which is consistent with previous work on hydrogen--water immiscibility \citep[e.g.,][]{Bali2013, Bailey2021, Vlasov2023, Bergermann2024, Cano2024, Gupta2025}. This divergence in the physically allowed range for $P_{1-2}$ could directly influence the inferred mass distribution and, consequently, the metallicities required to satisfy the observational constraints.

\section{Corner plots}\label{secA1}

\newpage

\newpage

\newpage

\newpage

\newpage

\newpage

\newpage

\end{appendix}

\end{document}